\documentclass[letterpaper]{sig-alternate-05-2015}
\usepackage{amsfonts}
\usepackage[colorlinks=true, citecolor=blue, urlcolor=cyan, linkcolor=red]{hyperref}
\usepackage{amssymb}
\usepackage{amsmath,amssymb,amscd,epsfig,amsfonts,rotating}
\usepackage{graphicx}
\usepackage{epstopdf}
\usepackage{subfigure}
\usepackage{multirow}
\usepackage{booktabs}
\usepackage{color,xcolor}
\usepackage{url}
\usepackage{amsmath,amscd,amsbsy,amssymb,latexsym,url,bm}
\usepackage{epsfig,epstopdf,graphicx,subfigure}
\usepackage{enumitem,balance}
\usepackage{wrapfig}
\usepackage{mathrsfs, euscript}
\usepackage{algorithm}
\usepackage{algorithmic}
\usepackage{amssymb}
\usepackage{ifpdf}
\usepackage{diagbox}
\usepackage{caption}
\usepackage{makecell}

\def\yoyi{YOYI}

\newfont{\mycrnotice}{ptmr8t at 7pt}
\newfont{\myconfname}{ptmri8t at 7pt}
\let\crnotice\mycrnotice%
\let\confname\myconfname%

\newtheorem{def:def}{Definition}
\newtheorem{thm:thm}{Theorem}
\newtheorem{thm:lm}{Lemma}
\newcommand{\bs}{\boldsymbol}

\DeclareMathOperator*{\argmax}{arg \, max}

\DeclareMathOperator*{\var}{\text{Var}}
\DeclareMathOperator*{\std}{\text{Std}}

\usepackage[margin=0.80in]{geometry}
\usepackage{array}

\begin{document}

\setcopyright{acmcopyright}

\CopyrightYear{2017}
\setcopyright{acmcopyright}
\conferenceinfo{WSDM 2017,}{February 06-10, 2017, Cambridge, United Kingdom}
\isbn{978-1-4503-4675-7/17/02}\acmPrice{\$15.00}
\doi{http://dx.doi.org/10.1145/3018661.3018701}
\clubpenalty=10000
\widowpenalty = 10000

\title{Managing Risk of Bidding in Display Advertising}
\author{
\alignauthor
Haifeng Zhang$^\dag$, Weinan Zhang$^{\ddag}$\thanks{Weinan Zhang is the corresponding author of this paper.}, Yifei Rong$^\sharp$, Kan Ren$^\ddag$, Wenxin Li$^\dag$, Jun Wang$^\natural$\\
      \affaddr{$^\dag$Peking University, $^\ddag$Shanghai Jiao Tong University,\\$^\sharp$YOYI Inc., $^\natural$University College London}\\
       \email{pkuzhf@pku.edu.cn, wnzhang@sjtu.edu.cn, yifei.rong@yoyi.com.cn}
}

\maketitle

\begin{abstract}
In this paper, we deal with the uncertainty of bidding for display advertising. Similar to the financial market trading, real-time bidding (RTB) based display advertising employs an auction mechanism to automate the impression level media buying; and running a campaign is no different than an investment of acquiring new customers in return for obtaining additional converted sales. Thus, how to optimally bid on an ad impression to drive the profit and return-on-investment becomes essential. However, the large randomness of the user behaviors and the cost uncertainty caused by the auction competition may result in a significant risk from the campaign performance estimation.
In this paper, we explicitly model the uncertainty of user click-through rate estimation and auction competition to capture the risk. We borrow an idea from finance and derive the value at risk for each ad display opportunity. Our formulation results in two risk-aware bidding strategies that penalize risky ad impressions and focus more on the ones with higher expected return and lower risk. The empirical study on real-world data demonstrates the effectiveness of our proposed risk-aware bidding strategies: yielding profit gains of 15.4\% in offline experiments and up to 17.5\% in an online A/B test on a commercial RTB platform over the widely applied bidding strategies.
\end{abstract}


\section{Introduction}\label{sec:intro}
Display advertising has become a significant battlefield for big data \cite{wang2016display}. As the advertising transactions are aggregated across websites in real time, the display advertising industry has a unique opportunity to understand the internet traffic, user behaviors, and online transactions. In this paper, we have used datasets from a leading DSP (Demand Side Platform) in China, iPinYou, which processes up to 18 billion ad impressions per day \cite{ipinyou-Statistics}. The advertising platform \yoyi, which has deployed our proposed algorithms in this paper, currently handles 5 billion ad transactions daily. Moreover, Fikisu DSP claims to process 32 billion ad impressions daily \cite{Fikisu-Statistics}; Turn reports to handle 2.5 million per second in the peak time \cite{shen20150}. To get the picture of the scale, the New York Stock Exchange trades around 12 billion shares daily \cite{NYSE-Statistics}, while the Shanghai Stock Exchange trades about 14 billion shares daily \cite{SSE-Statistics}. It is fair to say that the transaction volume from display advertising has already surpassed that of the financial market.

To further compare, similar to the financial market, ad impression trades are also automated largely by auction mechanisms --- while the financial market uses a double auction to create bid and ask quotes \cite{avellaneda2008high}, RTB (Real-time Bidding) display advertising adopts the second-price auction to gather bid quotes from advertisers once an impression is being generated \cite{yuan2013real,muthukrishnan2009ad}. As it is possible now to track user actions resulted from an online campaign, advertising optimization becomes more resembling to that of the financial market trading and tends to be driven by the marketing profit and return-on-investment (ROI). That is, there is explicit and measurable campaign goal of acquiring new users from the campaign in order to obtain additional sales from the acquired users. Thus, how to properly bid an ad impression to drive the profit and ROI becomes essential to performance-driven campaigns \cite{perlich2012bid,zhang2015statistical}.

Bidding strategies are normally built by estimating the utility of each ad impression, which is commonly done by predicting the underlying user's click-through rate (CTR) or conversion rate (CVR) \cite{lee2012estimating}. Existing solutions for predicting CTR range from linear models \cite{graepel2010web,mcmahan2013ad,he2014practical,lee2012estimating,richardson2007predicting} and gradient boosting decision trees (GBDT) \cite{trofimov2012using} to factorization machines \cite{oentaryo2014predicting}. All of them aim to return a predicted CTR value of the given ad impression. However, we never know whether such a \emph{point} estimation is confident enough. The true underlying CTR may heavily deviate from the predicted value, given the significant uncertainty of the underlying user behavior.
Such utility fluctuation, along with the cost uncertainty caused by the auction competitions \cite{amin2012budget}, results in a significant risk of the campaign profit estimation which should not be ignored.

In this paper, we depart from conventional CTR point estimations and explicitly model the CTR distribution to capture the uncertainty (risk) of the utility measure for each potential ad impression. Our idea is inspired from finance about \emph{value at risk}: a statistical technique used to measure and quantify the level of financial risk with an \text{investment} \cite{linsmeier2000value}.
The value at risk of an ad display opportunity is generally defined as the lower bound of the ad impression value guaranteed by a specific probability (risk). With the measured CTR prediction and market price risks, we propose two practical risk-aware bidding strategies to handle the uncertainty from both utility estimation and market competition.
Our methods are evaluated on two large-scale real-world ad datasets. With campaign profit as the key performance indicator (KPI), we find that our risk-averse bidding strategies, which penalize the bids with high uncertain CTR (or profit) and award the confident ones, yield a 15.4\% profit gain over a linear bidding strategy with a conventional logistic regression \text{CTR} estimator \cite{perlich2012bid} and largely outperform the risk-neutral and risk-seeking bidding.

Furthermore, we have deployed the risk-averse bidding strategies on \yoyi's DSP and conducted a 7-day online A/B test. In the live test, our proposed risk-aware bidding strategies bring 17.5\% higher campaign profit and 61.5\% higher CTR over the conventional ones by effectively saving money on uncertain and low-value opportunities, which verifies the practical effectiveness of our solutions of managing advertising bidding risk.

The rest of this paper is organized as follows. We discuss the related work in Section~\ref{sec:related-work}. Our CTR distribution modeling is provided in Section~\ref{sec:ctr-model}. In Section~\ref{sec:bid-model}, the concept of value at risk and the derived risk-aware bidding strategies are discussed. Offline and online experimental results are provided in Sections~\ref{sec:exp} and \ref{sec:online}, respectively. We finally conclude this paper and discuss our future work in Section~\ref{sec:con}.

\section{Related Work}\label{sec:related-work}



\noindent \textbf{User Response Prediction.}
Predicting the probability of a specific user response, e.g., CTR and CVR, is a key function for performance-driven online advertising \cite{graepel2010web,mcmahan2013ad,he2014practical}.
The applied CTR estimation models today are mostly linear. Logistic regression is the most widely used model, normally trained by stochastic gradient descent (SGD) \cite{lee2012estimating,richardson2007predicting}. The authors in \cite{mcmahan2013ad} proposed to use an online learning algorithm called follow-the-regularized-leader (FTRL) to train logistic regression from the streaming data. The model successfully bypasses the learning rate update problem in SGD and it empirically works effectively. Bayesian probit regression \cite{graepel2010web} is another linear model for online learning where the feature weights are modeled with a distribution and the model learning is via updating the weight posterior. Binary naive Bayes \cite{hand2001idiot} is also a popular linear model, by assuming the features are conditionally independent.

Linear models are simple and effective in learning, but may fail to capture the interactions between the assumed (conditionally) independent raw features \cite{graepel2010web}. By contrast, non-linear models are capable of learning feature interactions in various ways and could potentially improve prediction performance \cite{oentaryo2014predicting,trofimov2012using}. Gradient boosting decision trees (GBDT) \cite{trofimov2012using,he2014practical} are a straightforward non-linear model to capture feature interactions. Moreover, latent factor models, particularly factorization machines (FMs) \cite{oentaryo2014predicting}, map each binary feature into a low dimensional continuous space, and the feature interaction is automatically explored via vector inner product.

\vspace{5pt} \noindent \textbf{Real-Time Bidding Strategies.}
The emergence of ad exchanges for display advertising in 2009 \cite{muthukrishnan2009ad} provides automatic trading mechanism for advertisers to buy media inventory in impression level and determine the acceptable price via second price auction \cite{yuan2013real}.

The authors in \cite{ghosh2009adaptive} proposed an algorithm that made dynamic bidding decisions to achieve an optimal delivery with the budget constraint.
In \cite{chen2011real}, the bid price from each campaign was adjusted by the publisher or the supply-side platform in real time and the goal was to maximize the publisher side profit.
Borrowing the idea of the optimal truth-telling bidding in sponsored search \cite{edelman2005internet}, a basic bidding strategy is to bid the estimated true value for each ad impression. For performance-driven campaigns, the predefined true value is normally based on the economical value of the user actions, such as a click or a conversion. The expected true value for a specific impression is estimated as action value multiplied by action rate \cite{lee2012estimating,chen2011real}.
However, the truth-telling bidding strategy is optimal only when the budget and auction volume are not considered. With the campaign budget and lifetime auction volume constraints, the optimal bidding strategy may not be truth-telling. Extending from the truth-telling bidding strategy, the authors in \cite{perlich2012bid} proposed the generalized bidding function with a linear relationship to the predicted CTR for each ad impression being auctioned. Compared to \cite{perlich2012bid}, the authors in \cite{zhang2014optimal} proposed a functional optimization framework to directly optimize the bidding function, where the derived function showed that an optimal bidding function could be non-linear w.r.t. predicted CTR. The non-linearity is closely related to the distribution of market price \cite{amin2012budget}. Extending to multiple campaign bid optimization task, the authors in \cite{zhang2015statistical} further devised linear and non-linear bidding functions from the functional optimization framework. Extending from previous strategies, our work introduces a new factor, i.e. the standard deviation of predicted CTR, to be considered by bidding function in addition to predicted CTR.

\vspace{5pt} \noindent \textbf{Risk Management and Applications.}
Risk is a consequence of action taken in spite of uncertainty \cite{mun2006modeling}. The objective of risk management is to assure uncertainty does not to-some-extent deflect the business from its goals \cite{crouhy2006essentials}. 

Modern portfolio theory (MPT) \cite{markowitz1952portfolio} originates from modeling uncertainty of the return of combinations of multiple financial assets. It presents a quantitative method to measure such uncertainty (or risk) and embeds it into the decision making of investment \cite{hull2012risk}. In MPT, the variance of the return of each asset is modeled as its risk. Then the risk and expected return of a portfolio of invested assets are quantified by the fund allocation, the mean return of the assets and their covariance matrix \cite{markowitz1952portfolio,hull2012risk}.
MPT utilizes the mean-variance analysis to make an investment portfolio for any tradeoff between the risk and the expected return, or w.r.t. a reference investment such as bank rates \cite{sharpe1998sharpe}. With such advantages, MPT has been adopted in almost everywhere of financial investment \cite{elton2009modern}.

Recently, the ideas of risk management has been introduced to information retrieval, such as document ranking in web search \cite{wang2009portfolio,wang2012robust} and diversification in top-N recommendation \cite{shi2012adaptive,zhu2014mobile}, to improve the model robustness or catch the users' satisfaction on uncertainty psychologically. In the area of recommender system, bandits solutions \cite{zhao2013interactive} have been proposed to model confidence interval to balance exploration and exploitation in a risk-seeking fashion.

Computational advertising is associated with a certain level of deficit risk, particularly for performance-driven campaigns as the goal is to acquire new users and gain more sales from them. The risk comes from the dynamics of the market and the user online behaviors \cite{wang2012selling}. The authors in \cite{zhang2015statistical} proposed to measure campaign-level risk and return in a special case of arbitrage between CPM and CPA. Compared to \cite{zhang2015statistical}, our work focuses on single campaign optimization, and our risk is modeled from the uncertainty of user response and market competition at impression-level. Generally, our work borrows the concept of value at risk from finance to derive risk-aware bidding strategies intending to reasonably allocate budget between uncertain impressions and confident impressions and achieve a campaign-level profit gain, which is unlike in finance where risk control is only for balancing return and risk at item-level (impression-level in RTB).

\begin{table}
\small
\renewcommand{\arraystretch}{1}
\centering
\caption{Notations and descriptions}\label{tab:notation}
\vspace{-5pt}
\resizebox{\columnwidth}{!}{
\begin{tabular}{c|l}
\hline
Notation & Description \\
\hline
$\bs{x}$ & The features of bid request. \\
$\bs{w}$ & The weights of features in CTR estimation function. \\
$y$ & The true label of user response. \\
$\hat{y}$ & The estimated label of user response. \\
$\bs{\mu}$ & The mean vector of $\bs{w}$. \\
$\bs{S}$ & The covariance matrix of $\bs{w}$. \\
$q_i$ & The precision (reciprocal of variance) of $w_i$. \\
$v$ & The pre-defined value of positive user response. \\
$b$ & The bid price determined by bidding strategy. \\
$z$ & The market price determined by second-price auction. \\
$\alpha$ & The coefficient balancing utility and risk. \\
$R$ & The utility of advertiser. \\
$p()$ & The probability density function. \\
$P()$ & The probability mass function. \\
$\sigma()$ & The sigmoid function.\\
\hline
\end{tabular}
}
\end{table}

\section{CTR Distribution Modeling}\label{sec:ctr-model}

RTB is generally a two-phase process: 1) CTR estimation based on features of bid request; 2) bid price determination based on estimated CTR. In this section, we explicitly model the CTR distribution in order to deal with the uncertainty of a CTR estimation. Next, in Section~\ref{sec:bid-model} we shall propose risk-aware bidding strategies from the inferred CTR distribution and the market price distribution.

In this paper, we don't provide down-to-earth background of RTB. For more details about RTB, we refer to \cite{wang2016display}. For clarity, we summarise notations in this paper in Table~\ref{tab:notation}.

\subsection{\hbox{Preliminary: Bayesian Logistic Regression}} \label{sec:bayes-lr}
We propose to use a Bayesian logistic regression to model the CTR distribution due to the following reasons: (i) logistic regression (LR) has been widely deployed as the CTR prediction model in most RTB ad platforms \cite{lee2012estimating,perlich2012bid} and our model is a natural extension to tackle the uncertainty of a CTR estimation with LR; (ii) We adopt Bayesian treatment to model uncertainty since it has been well studied by previous works \cite{zhang2010learning,graepel2010web} for CTR estimation; (iii) Although Bayesian probit regression \cite{zhang2010learning,graepel2010web} has the potential to model the uncertainty, its probit activation function is of no closed form, and thus is computationally low cost effectiveness in RTB.


For readability, in this section, we present a preliminary on Bayesian logistic regression, while for details, we refer to \cite{bishop2006pattern}.  For a multi-dimensional feature vector $\bs{x}$ representing the input ad display opportunity, the conventional logistic regression estimates the CTR by:
\begin{align}
\hat{y} = \sigma(\bs{w}^T\bs{x}) = \frac{1}{1+e^{-\bs{w}^T\bs{x}}} \label{eq:lr},
\end{align}
where $\sigma$ is the sigmoid function and $\bs{w}$ is the weight vector of logistic regression. The likelihood of observing the correct binary click label $y$ given features $\bs{x}$ and weights $\bs{w}$ is
\begin{align}
p(y|\bs{x}, \bs{w}) = \sigma(\bs{w}^T\bs{x})^{y}(1-\sigma(\bs{w}^T\bs{x}))^{(1-y)}.
\end{align}

In the Bayesian version of logistic regression, $\bs{w}$ is modeled as a random variable with a p.d.f. $p(\bs{w})$. Thus the marginal conditional probability $p(y|\bs{x})$ is
\begin{align}
p(y|\bs{x}) = \int_{\bs{w}} p(y|\bs{x}, \bs{w}) p(\bs{w}) d\bs{w}.
\end{align}

We follow \cite{zhang2010learning,graepel2010web} to adopt a Gaussian prior $N(\bs{\mu_0}, q_0^{-1}\bs{I})$ on $\bs{w}$, which is a practical setting. After observing a data instance $(\bs{x}, y)$, the posterior distribution of $\bs{w}$ becomes
\begin{align}
& p(\bs{w}|\bs{x},y) = \frac{p(y|\bs{x}, \bs{w})p(\bs{w})}{\int_{\bs{w}'}p(y|\bs{x}, \bs{w}')p(\bs{w}') d\bs{w}' }  \label{eq:intractable-post}\\
\propto & \sigma(\bs{w}^T\bs{x})^{y}(1-\sigma(\bs{w}^T\bs{x}))^{(1-y)}   \prod_i \sqrt{\frac{q_{\text{old},i}}{2\pi}}e^{-\frac{(w_i-\mu_{\text{old},i})^2 q_{\text{old},i}}{2}}, \nonumber
\end{align}
where $\mu_{\text{old},i}$ and $q_{\text{old},i}$ are the prior parameters of the $i$-th dimension of $\bs{w}$ before observing the data instance $(\bs{x}, y)$.

\vspace{5pt} \noindent \textbf{Approximation of Posterior.}
Since the posterior Eq.~(\ref{eq:intractable-post}) is computationally complex, we maintain a Laplace approximation \cite{bishop2006pattern} to keep it consistent with the prior. There are alternative approximate inferences such as variational inference (VI) \cite{bishop2006pattern}. In this paper we adopt Laplace approximation due to its simpler implementation and lower computational cost than VI, which involves extra variational parameters and invokes a time-consuming EM algorithm for training. The Gaussian approximation to the posterior distribution takes the form
\begin{align}
p(\bs{w})=N(\bs{w}|\bs{\mu}_{\text{new}},\bs{S}_{\text{new}}),
\end{align}
where $\bs{\mu}_{\text{new}}$ is defined by the $\bs{w}_{\text{MAP}}$ which maximizes the logarithmic posterior:
\begin{align}
\bs{\mu_{\text{new}}} = & \argmax_{\bs{w}} y\ln \sigma(\bs{w}^T\bs{x}) + (1-y)\ln (1-\sigma(\bs{w}^T\bs{x})) \nonumber \\
& ~~~~~~~~~~ - \frac{1}{2} \sum_i q_{\text{old},i} (w_i-\mu_{\text{old},i})^2 + \text{const} , \label{eq:update-mu}
\end{align}

whose SGD updating is
\begin{align}
\mu_{\text{new}, i} \leftarrow & \mu_{\text{new}, i} + \eta \cdot \Big((y - \sigma(\bs{\mu}_{\text{new}}^T\bs{x}))x_i \nonumber \\
& ~~~~~~~~~~~~~~~~~~~ - q_{\text{old},i} (\mu_{\text{new},i} - \mu_{\text{old},i})\Big)
\end{align}
and $\bs{S}_{\text{new}}$ is given by the inverse of the matrix of second derivatives of the negative log likelihood, which satisfies
\begin{align}
\bs{S}_{\text{new}}^{-1} = & -\nabla \nabla \ln p(\bs{\mu}|\bs{x}, y) = \bs{S}_{\text{old}}^{-1} + \sigma(\bs{\mu}^T\bs{x})(1-\sigma(\bs{\mu}^T\bs{x}))\bs{x}\bs{x}^T. \nonumber
\end{align}

As we follow \cite{zhang2010learning,graepel2010web} to assume each $w_i$ as independent with each other and thus $\bs{S}$ is diagonal, we have the updating of precision parameters as
\begin{align}
q_{\text{new},i} = \bs{S}_{\text{new},i,i}^{-1} = q_{\text{old},i} + \sigma(\bs{\mu}^T\bs{x})(1-\sigma(\bs{\mu}^T\bs{x}))x_i^2. \label{eq:update-q}
\end{align}



\subsection{Predicted CTR Distribution} \label{sec:pred-ctr-pdf}
Equipped with a Bayesian logistic regression with a Laplace approximation of the parameter posterior, we are ready to propose a simple yet novel solution for modeling the CTR distribution. Note that the CTR itself is a probability estimation that a click event occurs from an impression with feature $\bs{x}$ takes the form
\begin{align}
\hat{y} = P(y=1|\bs{x}) = \sigma(\bs{w}^T\bs{x}),~~~w_i \sim N(\mu_i, q_i^{-1}),
\end{align}
where $\hat{y}$ denotes the CTR random variable that generates the binary observation $y$, so our goal is to estimate the distribution of $\hat{y}$. As introduced in Section~\ref{sec:bayes-lr}, we assume each $w_i$ is from Gaussian i.i.d., the distribution of $\sum_i w_i x_i$ also follows $N(\sum_i \mu_i x_i, \sum_i q_i^{-1} x_i)$. Thus, we have
\begin{align}
\hat{y} = \sigma(a),~~~a \sim N\Big(\sum_i \mu_i x_i, \sum_i q_i^{-1} x_i \Big)~.
\end{align}

Consider that if random variables $x$ and $y$ satisfy $y = g(x)$ and $g^{-1}$ is monotonic and differentiable, we have \cite{wasserman2013all}
\begin{align}
p_y(y) = p_x(g^{-1}(y))\Big|\frac{\partial g^{-1}(y)}{\partial y} \Big|~.
\end{align}

\begin{figure}[t]
  \centering
  \vspace{-8pt}
  \includegraphics[width=.7\columnwidth]{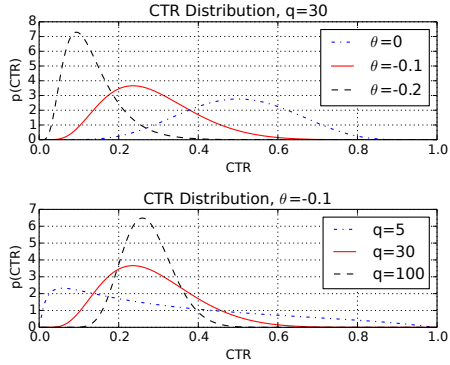}\\
  \vspace{-8pt}
  \caption{An illustration of the proposed CTR distribution against its parameters $\mu$ and $q$ in Eq.~(\ref{eq:ctr-pdf}).}\label{fig:ctr-pdf}
\end{figure}

In our case, $\sigma^{-1}(\hat{y}) = \ln \hat{y} - \ln (1-\hat{y})$ is monotonic and differentiable within $(0, 1)$, so we obtain the closed-form of $\hat{y}$'s p.d.f.
\begin{align}
p_{\hat{y}}(\hat{y}) = & \frac{1}{(\hat{y} - \hat{y}^2)\sqrt{2\pi \sum_i q_i^{-1} x_i}}e^{-\frac{(\sigma^{-1}(\hat{y})-\sum_i \mu_i x_i)^2 }{2 \sum_i q_i^{-1} x_i}}, \label{eq:ctr-pdf}
\end{align}
which provides an explicit CTR p.d.f. To our best knowledge, the above proposed solution has not been studied in previous literature \cite{zhang2010learning,graepel2010web,wang2010click}.
To understand the nature of the model, Figure~\ref{fig:ctr-pdf} plots the CTR distribution against its parameters $\mu$ and $q$, where $\bs{x}=\bs{1}_{10}$. As observed, the p.d.f. presents a single-peak shape in $[0,1]$ which is similar with the shapes of beta distributions.

It is straightforward to see from Figure~\ref{fig:ctr-pdf} that $\mu$ and $q$ jointly determine the peak location and sharpness of the CTR p.d.f. and specifically $\mu$ influences more on the peak location while $q$ influences more on the distribution sharpness. To understand how it models the CTR prediction uncertainty/confidence, from Eq.~(\ref{eq:update-q}), we see each time a data instance with feature $\bs{x}_i$ is observed, the precision $q_i$ will be updated with a higher value, which in turn contributes a sharper CTR p.d.f. in Eq.~(\ref{eq:ctr-pdf}). Therefore, for the ad impression with frequent (similar) features, the predicted CTR is of low uncertainty, and vice versa.


\section{Risk-aware Bidding Strategies}\label{sec:bid-model}
With our CTR distribution model in Eq.~(\ref{eq:ctr-pdf}), we next investigate the conditional distribution of a utility given a specific bid price for an input bid request. By considering a risk-aware utility as an optimization target, we are ready  to derive the corresponding risk-aware bidding strategy. Note that alternative CTR distribution models can also be incorporated in our solution framework.

Specifically, we start from the theoretic derivation of a Bayesian truth-telling bidding strategy and provide an analysis of its risk in Section~\ref{sec:truth-telling}. Then we will propose two solutions, discussed in Sections~\ref{sec:bid-var} and \ref{sec:bid-rev-risk} respectively.

\subsection{Analysis: Bayesian Truth-telling Bidding}\label{sec:truth-telling}
The utility $r$ of an ad impression could be defined based on the advertiser's value $v$ on a specific user action, e.g., click or conversion. For example, if the value $v$ is on each click, given an impression with CTR $\hat{y}$ distributed as in Eq.~(\ref{eq:ctr-pdf}), the utility $r$ and its p.d.f. of this impression are
\begin{align}
r = v \cdot \hat{y}, ~~~ p_r(r) = p_{\hat{y}}(\hat{y}) / v . \label{eq:utility}
\end{align}

Moreover, the cost to win the ad impression comes from the highest bid from other competitors, defined as market price $z$ \cite{amin2012budget}.
The profit  of winning this impression is then calculated as the utility $r$ minus the cost $z$, which is set as the optimization target of performance-driven campaigns~\cite{zhang2015statistical}.


In a general setting, considering both the estimated CTR $\hat{y}$ and cost $z$ are stochastic variables: $\hat{y} \sim p_{\hat{y}}(\hat{y})$ and $z \sim p_z(z)$, the bid optimization problem is to find the optimal bid price to the auction.

Let us first consider a simple case without considering the uncertainty of our estimation, where the goal is to maximize the expected profit $R(b)$ by marginalizing out $\hat{y}$ and $z$:
\begin{align}
b^* &= \argmax_{b} \mathbb{E}[R(b)] \\
&= \argmax_{b} \int_{\hat{y}} \int_{z=0}^b (v \cdot \hat{y} - z) p_z(z) dz \cdot p_{\hat{y}}(\hat{y}) d \hat{y}. \label{eq:max-net-profit}
\end{align}
Take the derivative w.r.t. $b$ and set it to 0:
\begin{align}
\frac{\partial \mathbb{E}[R(b)]}{\partial b} &= \frac{\partial}{\partial b}  \int_{z=0}^b p_z(z) \int_{\hat{y}} (v \cdot \hat{y} - z) p_{\hat{y}}(\hat{y}) d\hat{y} dz \\
&= p_z(b) \int_{\hat{y}} (v \cdot \hat{y} - b) p_{\hat{y}}(\hat{y}) d\hat{y} = 0\\
\Rightarrow b^* &= \int_{\hat{y}} v \cdot \hat{y} \cdot p_{\hat{y}}(\hat{y}) d\hat{y} = v \cdot \mathbb{E}[\hat{y}] = \mathbb{E}[r]. \label{eq:true-bidding}
\end{align}
We see that the optimal bid price is the product of the action value and the estimated CTR $\hat{y}$, which is independent of the market price distribution. If we assume $\hat{y}$ is known and fixed, i.e., $p_{\hat{y}}(\hat{y})$ focuses its mass on a single point, then the optimal bid price is $v \cdot \hat{y}$. Note that the optimality of truth-telling bidding is for one-shot auction. When considering campaign budget and auction volume, a coefficient $\phi$ is commonly added to Eq.~(\ref{eq:true-bidding}) as $\phi \cdot \mathbb{E}[r]$.



\begin{figure}[t]
  \centering
  \vspace{-3pt}
  \includegraphics[width=.95\columnwidth]{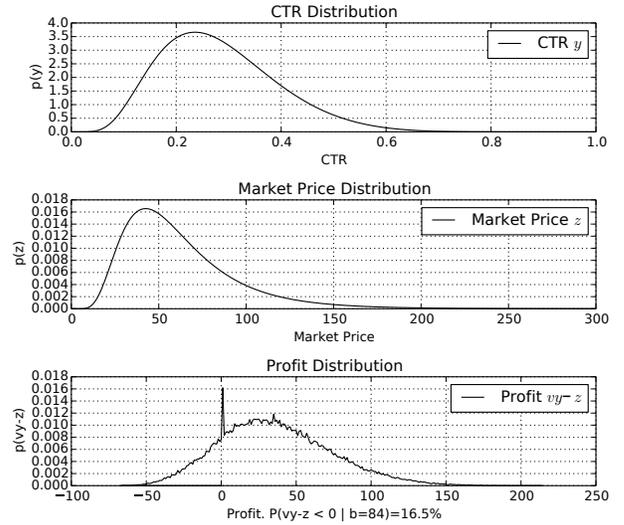}\\
  \vspace{-3pt}
  \caption{An example of CTR, market price and profit distribution when bidding the expected utility value. The profit p.d.f. on 0 is the probability of losing the auction, resulting in a peak.}\label{fig:ctr-mp-rev-pdf}
\end{figure}

\vspace{5pt}\noindent \textbf{Discussion of Risk.} The above classic bidding solution is built on maximizing the expectation of the profit $R(b)$, regardless of its uncertainty. However, a potential problem is that there is a chance a bid is won, but $v \cdot \hat{y}$ is less than $z$, in which case $R(b)$ will be negative. We can obtain the probability of such negative profit $P(R(b) < 0)$ as
\begin{align}
P(R(b) < 0) = & \int_{0}^{1} p_{\hat{y}}(\hat{y}) P(b > z > v \cdot \hat{y}) d\hat{y} \nonumber \\
= & \int_{0}^{1} p_{\hat{y}}(\hat{y}) \int_{v \cdot \hat{y}}^{b} p_z(z) dz d\hat{y}, \label{eq:negative-probability}
\end{align}
which shows it is of risk to get negative profit whatever the positive bid price. We illustrate our point in Figure~\ref{fig:ctr-mp-rev-pdf} using an example. We set the CTR p.d.f. as Eq.~(\ref{eq:ctr-pdf}) with $\sum_i \mu_i x_i = -1$, $\sum_i q_i^{-1} x_i = \frac{1}{3}$; the market price distribution is assumed to be a log-normal p.d.f. with $\mu=4$, $\sigma=0.5$; the value per click is set as 300. The expected value of CTR is 0.283, and thus the truth-telling bidding in Eq.~(\ref{eq:true-bidding})  would give 84. From our simulation with 10,000 samples, we plot the profit distribution given the truth-telling bid 84 with probability 16.7\% the profit is negative. From Eq.~(\ref{eq:negative-probability}) we see that the down-side risk, the probability of the CTR lower than the mean $\hat{y}$, contributes more to the probability of negative profit; the higher standard deviation (or variance) is of $\hat{y}$, the higher probability of such negative profit may occur in the truth-telling bidding case.
Thus, such risk of negative profit should be carefully considered and incorporated into the bidding strategies to help campaigns achieve satisfactory performance with controlled risk.

\subsection{Bidding with Value-at-Risk Utility}\label{sec:bid-var}
The utility p.d.f. $p_r(r)$ is of high importance for risk modeling. From Figure~\ref{fig:ctr-mp-rev-pdf} we know that due to the uncertainty of utility, bidding with the utility expectation $\mathbb{E}[r]$ will introduce the risk of negative profit, which might be worse than no bid. Let us examine the downside(upside) risk.

\begin{thm:lm}[Cantelli's Inequality]
For a random variable $r$ with mean $\mu$ and standard deviation $\sigma$, the following inequalities hold:
\begin{align}
P(r < \mu - \alpha \sigma) < \frac{1}{1 + \alpha^2}~, \alpha > 0, \\
P(r > \mu - \alpha \sigma) < \frac{1}{1 + \alpha^2}~, \alpha < 0.
\end{align}
\label{lm:cantelli}
\end{thm:lm}
With Lemma~\ref{lm:cantelli} we can find a variant of \emph{value at risk} (VaR) \cite{linsmeier2000value,rockafellar2000optimization} for utility
\begin{align}
\tilde{r} = \mathbb{E}[r] - \alpha \std[r],\label{eq:value-at-risk}
\end{align}
which has a guarantee that for $\alpha>0$ the real utility is lower than $\tilde{r}$ with a probability smaller than $1/(1+\alpha^2)$; for $\alpha<0$ the real utility is higher than $\tilde{r}$ with a probability smaller than $1/(1+\alpha^2)$.

Using VaR $\tilde{r}$ as a new risk-aware utility, we set the truth-telling bid at $\tilde{r}$.
Taking Eq.~(\ref{eq:utility}) into Eq.~(\ref{eq:value-at-risk}) gives
\begin{align}
b_{\text{VaR}}(r) = \tilde{r} = v \cdot (\mathbb{E}[\hat{y}] - \alpha \std[\hat{y}]), \label{eq:bid-var}
\end{align}
where the p.d.f. of $\hat{y}$ is given in Eq.~(\ref{eq:ctr-pdf}).

When $\alpha > 0$, the bidding strategy is risk-averse, inversely when $\alpha<0$ it is risk-seeking; the traditional truth-telling bidding (Eq.~(\ref{eq:true-bidding})) is now a special case of $b_{\text{VaR}}(r)$ with $\alpha = 0$, called risk-neutral.

Note that we derive VaR strategy based on Lemma~\ref{lm:cantelli} rather than the distribution of $\hat{y}$  (Eq.~(\ref{eq:ctr-pdf})) directly for the following reasons:
(i) Lemma~\ref{lm:cantelli} is a general one in the sense that it only needs mean and standard deviation, so it can be applied to other CTR distribution models, especially those without analytical forms;
(ii) $\alpha$ can be simply set over all bid requests to make the computation more efficient. The uniform $\alpha$ can also be regarded as a risk control parameter.

Considering campaign budget and auction volume, a coefficient $\phi$ is commonly added to Eq.~(\ref{eq:bid-var}), which is similar to truth-telling bidding \cite{perlich2012bid}. This also applies to the strategy proposed in Section~\ref{sec:bid-rev-risk}. The detailed realization method and efficiency analysis of both strategies are provided in the appendix.

\subsection{Bidding for Risk Management of Profit}\label{sec:bid-rev-risk}
Another way to approaching risk-aware bidding strategies is to go back to the analysis of campaign profit $R(b)$, which is one of the key performance indicators (KPIs) in RTB.

Mathematically,
\begin{align}
R(b) = \begin{cases}
0 & b \leq z~~\text{(lose)}\\
v \cdot \hat{y} - z & b > z~~\text{(win)}
\end{cases}~. \label{eq:profit}
\end{align}

As both CTR $\hat{y}$ and market price $z$ are modeled as stochastic random variables with p.d.f. $p_{\hat{y}}(\hat{y})$ and $p_z(z)$, $R(b)$ can be naturally regarded as a dependent random variable.
Again, using Lemma~\ref{lm:cantelli}, the value-at-risk of profit
\begin{align}
\tilde{R}(b) = \mathbb{E}[R(b)] - \alpha \std[R(b)],\label{eq:value-at-risk-profit}
\end{align}
which has a guarantee that for $\alpha>0$ the real profit is lower than $\tilde{R}(b)$ with a probability smaller than $1/(1+\alpha^2)$; symmetrically for $\alpha<0$.

In this setting, we propose the second risk-aware bidding strategy that generates the bid which yields the maximum value-at-risk of profit:
\begin{align}
b_{\text{RMP}}(R(b)) = \argmax_b ~ \mathbb{E}[R(b)] - \alpha \std[R(b)]. \label{eq:bid-rmr}
\end{align}

Through analyzing the properties of $R(b)$, we find that $\mathbb{E}[R(b)]$ and $\std[R(b)]$ have a non-trivial trade-off relation, which can be balanced by $b$.
We solve the optimal $b$ by importing the concept of efficient frontier from finance \cite{hull2012risk}. Detailed analysis can be found in the appendix.

\section{Offline Empirical Study}\label{sec:exp}
In this section, we empirically evaluate our proposed risk-aware bidding strategies on real-world data\footnote{Our experiment is repeatable and the code is given below: \url{https://github.com/pkuzhf/bidding-at-risk-experiment}}.


\subsection{Datasets}
Two ad log datasets in our empirical study are:
\begin{description}
\vspace{-5pt}
\item[iPinYou] is the largest independent programmatic media buying platform in China. The dataset\footnote{Dataset link: \url{http://data.computational-advertising.org}} was released for research after its global competition on RTB algorithms in 2013 \cite{liao2014ipinyou}. It contains 19.50M ad impressions, 14.89K clicks from 9 campaigns during 10 days in 2013, which involve 16K Chinese Yuan (CNY) expense in total. According to the data publishers, the last three day data of each campaign is split into test data while the former part into training data. The overall effective cost per click (eCPC) is 1.07 CNY on the training data and 1.13 CNY on the test data, which means the overall market competition did not change dramatically during that 10 days.
\vspace{-5pt}
\item[\yoyi] is another mainstream demand-side platform company. This proprietary dataset is mainly used for training the risk-aware models deployed for online A/B testing. Details will be given in Section~\ref{sec:online}.
\end{description}

\vspace{-5pt}
Both datasets are with record-per-line format. After ad log joining, each record is formalized as a triple $(\bs{x}, y, z)$, where $\bs{x}$ is the high dimensional feature vector for each bid request with the corresponding ad information, $y$ is the user feedback on the ad impression, e.g., the binary click or conversion actions, $z$ is the market price for that auction, i.e., the lowest price to bid in order to win the auction.

\subsection{Experiment Protocol}
We followed \cite{zhang2015statistical} to set up our evaluation procedure. For each campaign bidding strategy with a predefined budget, the optimal parameters $(\bs{\mu},\bs{S})$ were learned on training data and the hyperparameters ($\alpha$ and $\phi$, discussed later) were tuned on validation data, which was the early half split from the test data. Then we replayed the historic bid records to test the performance on the other half of test data. We did not use cross-validation since our data is sequential.

For each campaign, there was a defined value $v$ for the user action, i.e., a click in our experiment. Following \cite{zhang2015statistical} the click value is defined by a proportion of eCPC on training data to imitate the true value set by the advertisers. In this paper, we set the proportion to $100\%$. Every time the tested bidding agent received a bid request from the ad exchange, it generated a bid with the tested bidding strategy. If the bid price was higher than the historic market price, the agent won the ad impression and paid the market price and then the historic recorded binary user feedback, i.e., click or not, was observed. If there was a user click, the agent made revenue of the click value. The test ended either when there was no more test bid request or when the campaign budget was exhausted, if applicable.

Note that the offline experiment cannot fully simulate the market competition because there is no observed user feedback and market price for historically lost auctions. Nevertheless, our evaluation protocol keeps the bid requests, displayed ads, and auction environment unchanged. We try to answer that under the same context if the campaign was given a different bidding strategy, whether they would be able to get more clicks with the budget limitation.

\subsection{Compared Bidding Strategies}\label{sec:compared-strategy}
The compared bidding strategies are as follows.
\begin{itemize}
\vspace{-5pt}
\item \textbf{LR} - Linear Revenue Bidding: the baseline, which is the most widely used bidding function
\begin{align}
b_\text{LR} = \phi \cdot v \cdot \hat{y},\label{eq:lr-bid}
\end{align}
where $\hat{y}$ is the predicted CTR with logistic regression as in Eq.~(\ref{eq:lr}). $\phi$ is the scaling parameter tuned based on the market competitiveness \cite{perlich2012bid}. The overall AUC of this LR estimator on iPinYou dataset is $69\%$.
\vspace{-5pt}
\item \textbf{VaR} - Value at Risk Bidding: our first risk-aware bidding strategy proposed in Section~\ref{sec:bid-var}, which bids the value at risk, shown in Eq.~(\ref{eq:bid-var}), where $\alpha$ is the hyperparameter to be tuned on validation data.
\vspace{-5pt}
\item \textbf{RMP} - Risk Management of Profit based Bidding: the second risk-aware bidding strategy proposed in Section~\ref{sec:bid-rev-risk}, which seeks the optimal bid to maximize the expected profit with its standard deviation as risk constraint, shown in Eq.~(\ref{eq:bid-rmr}), where $\alpha$ is the hyperparameter to be tuned on validation data.
\end{itemize}

The two risk-aware bidding strategies rely on the same CTR distribution model proposed in Section~\ref{sec:ctr-model}. The initial prior distribution of $\bs{w}$ was given that (i) $\bs{\mu}_0$ was set as the point estimation generated by logistic regression;
(ii) $q_0$ was set as constant 1. Our model achieves a comparable AUC with standard LR model. Note that for the budgeted bid optimization tasks in Section~\ref{sec:budget-exp}, there would also be a hyperparameter $\phi$ multiplied in VaR (\ref{eq:bid-var}) and RMP (\ref{eq:bid-rmr}) to be tuned to avoid budget under- or over-delivery.



\subsection{Evaluation Measures}\label{sec:eval-measure}
For a bid optimization task, the major evaluation measures are campaign profit, i.e., the earned total click value minus the cost, and ROI, i.e., the ratio between the profit and the cost. In addition, we also monitor other key metrics, such as cost per thousand impressions (CPM), auction winning rate, CTR and eCPC to gain more insights.

In order to investigate how bidding strategies balance return against risk, we also propose an additional metric: profit - $\lambda$ cost, which is named as Cost-Penalized Profit (CP-Profit) in this paper. Intuitively, advertisers want to maximize the profit of their performance-driven campaign given the budget, or minimize the advertising cost given the profit, either of which forms a profit/cost tradeoff. Note that this metric was mainly used for model selection, i.e., selecting $\alpha$ and $\phi$. As we will show in Section~\ref{sec:cp-profit} about the profit/cost analysis, CP-Profit is an effective metric to balance the profit and the cost to select an optimal model.

\subsection{Profit and Cost Analysis}\label{sec:cp-profit}
We analyzed the profit and cost of various bidding strategies with various parameter settings. This would help us understand the properties of the bidding strategies and help us select optimal model hyperparameters.

\begin{figure}[t]
 \centering
 \vspace{-1em}
  \subfigure{\includegraphics[width=0.49\columnwidth]{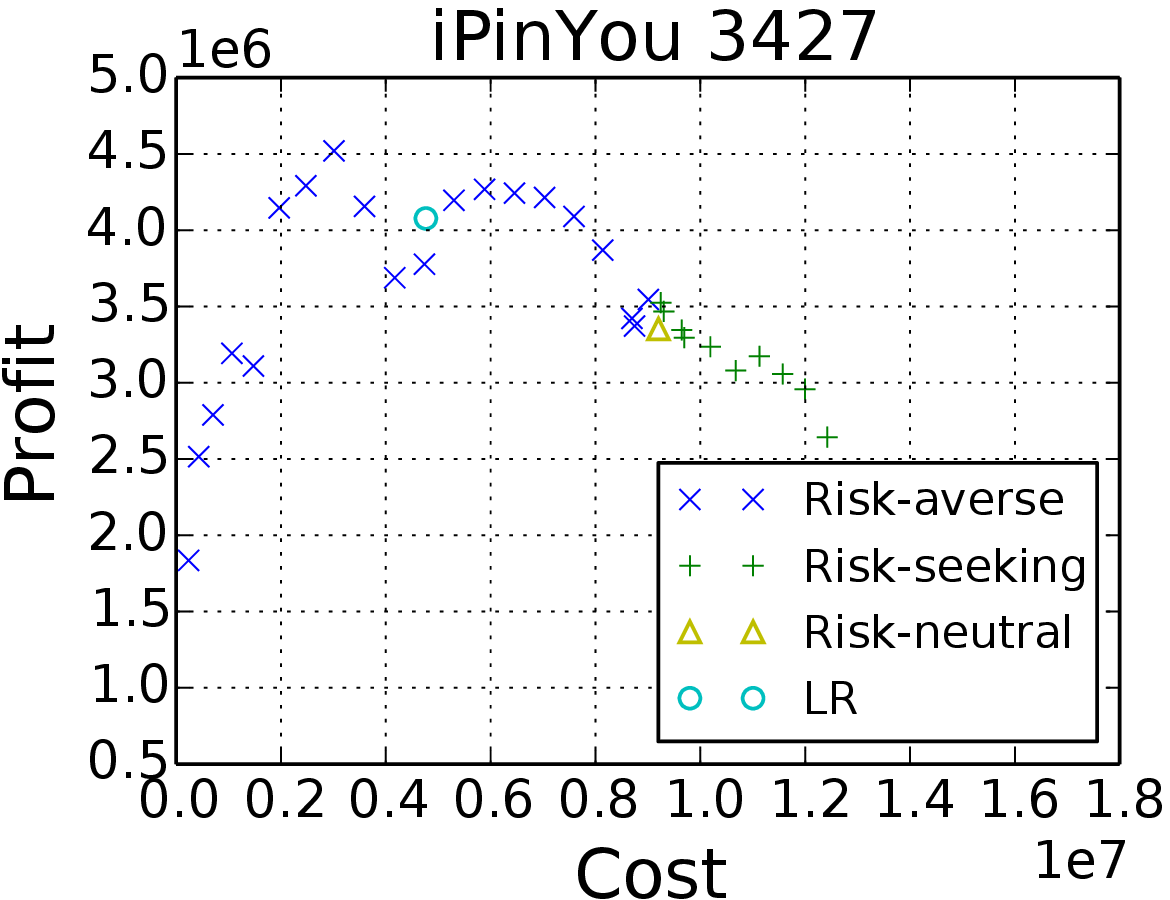}}
  \subfigure{\includegraphics[width=0.49\columnwidth]{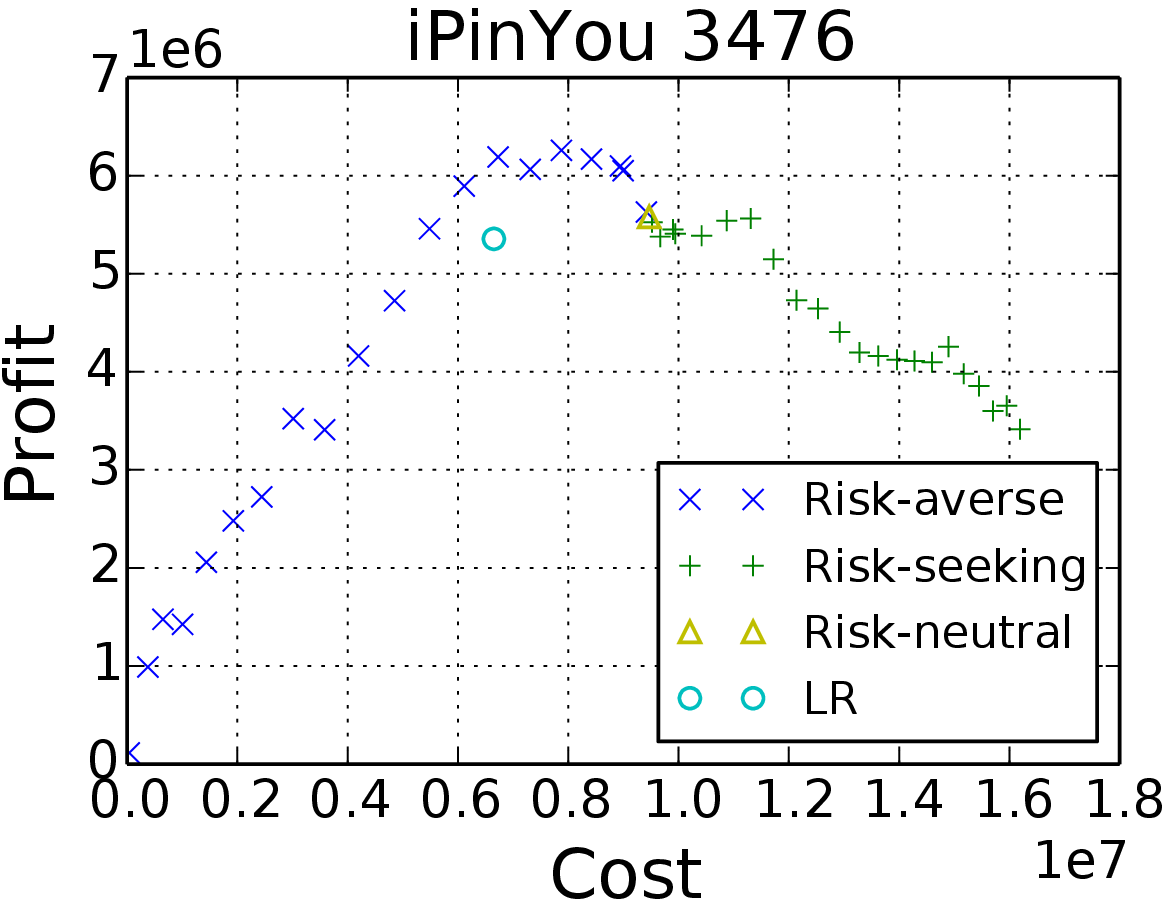}}
 \vspace{-8pt}
 \caption{Profit v.s. cost of the VaR bidding.}
 \label{fig:rev-cost-analysis-var}
\end{figure}

\begin{figure}[t]
 \centering
  \vspace{-1em}
  \subfigure{\includegraphics[width=0.49\columnwidth]{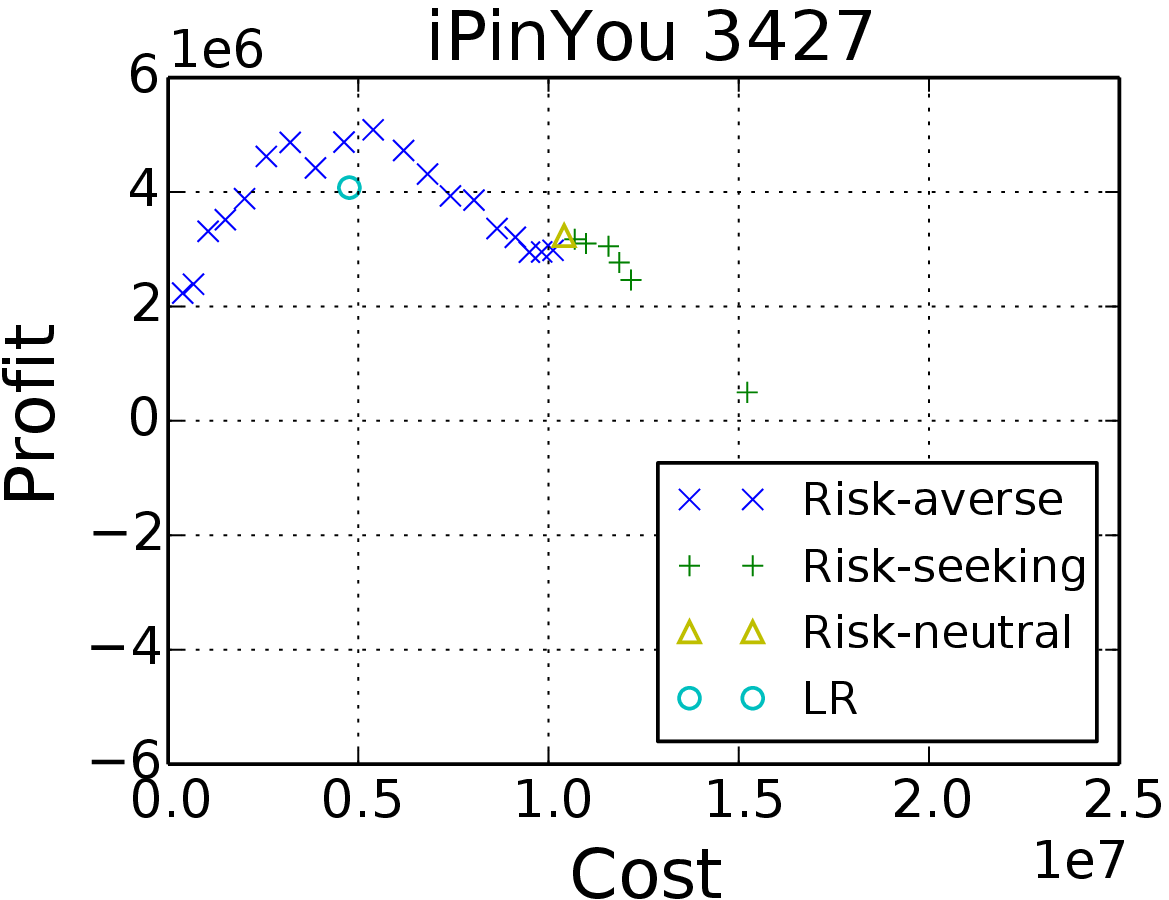}}
  \subfigure{\includegraphics[width=0.49\columnwidth]{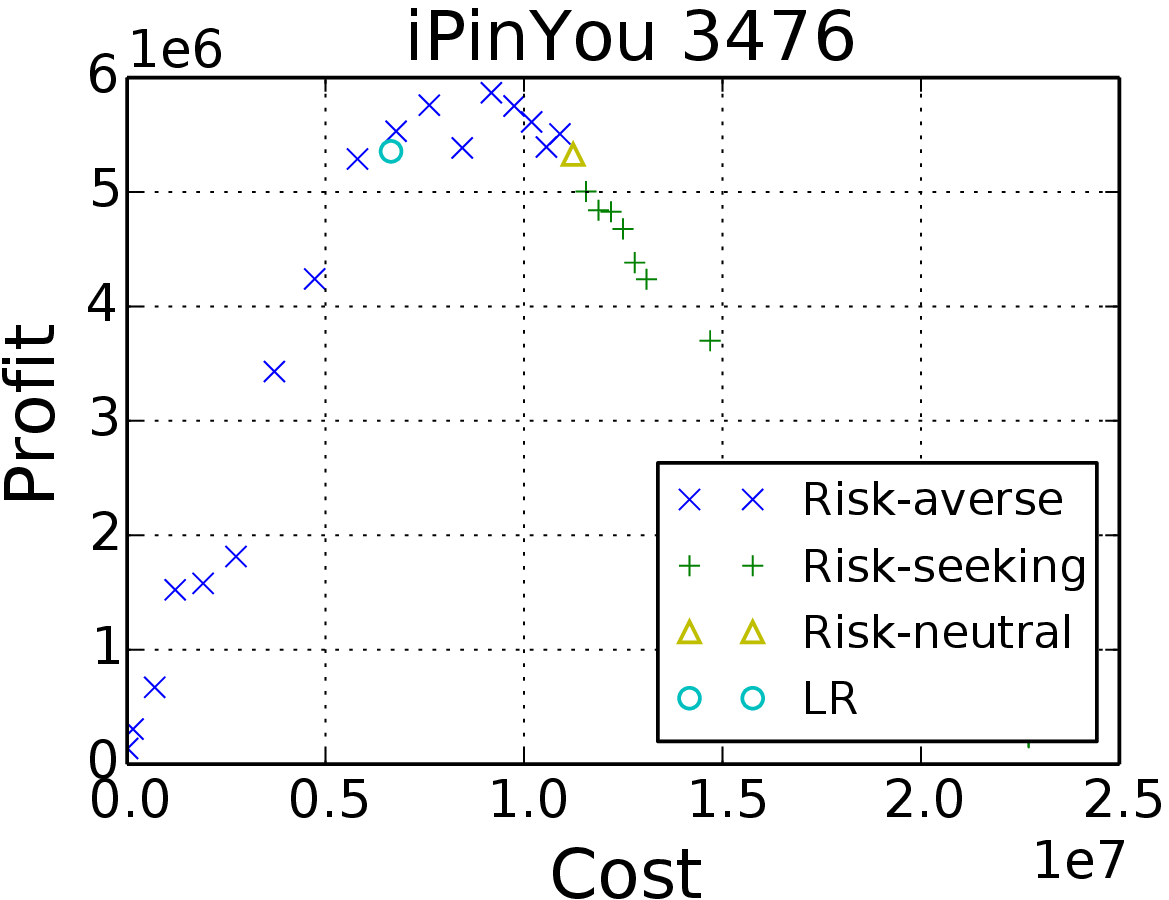}}
 \vspace{-8pt}
 \caption{Profit v.s. cost of the RMP bidding.}
 \label{fig:rev-cost-analysis-rmr}
\end{figure}

\vspace{5pt} \noindent \textbf{Non-Budgeted Settings.}
Without budget constraint, the basic LR bidding strategy is truth-telling, i.e. $\phi=1$. For the risk-aware bidding strategies we varied the value of $\alpha$. Figure~\ref{fig:rev-cost-analysis-var} shows the profit and cost change when setting different $\alpha$'s in VaR Eq.~(\ref{eq:bid-var}), with $\alpha>0$ as risk-averse, $\alpha<0$ as risk-seeking and $\alpha=0$ as risk-neutral. We clearly observed that for all tested campaigns the results generally formed a single-peak concave shape, showing strong trade-off between the profit and cost. Risk-averse strategies intended to yield lower cost than risk-seeking ones, as the risk-averse strategies would always bid lower than the risk-neutral one, and the latter further bid lower than the risk-seeking ones. The performance-driven campaigns pursue high profit and low cost. Thus if a bidding strategy yields higher profit and lower cost than another one, we say the former strategy \emph{dominates} the latter one. From the results we saw that almost all the risk-seeking strategies were dominated by part of the risk-averse strategies. That is to say, for almost every risk-seeking setting with $\alpha<0$, there always exists one or more risk-averse setting with $\alpha>0$ that yields equal or higher profit with lower cost.

Figure~\ref{fig:rev-cost-analysis-rmr} illustrates the profit and cost trade-off yielded by RMP Eq.~(\ref{eq:bid-rmr}) with various risk settings. We also observed that, similarly, the risk-averse strategies tended to yield lower cost than the risk-seeking ones. The risk-neutral strategy acted as a splitting point between the risk-averse and risk-seeking ones.

\begin{figure}[t]
  \centering
  \vspace{-8pt}
  \includegraphics[width=0.8\columnwidth]{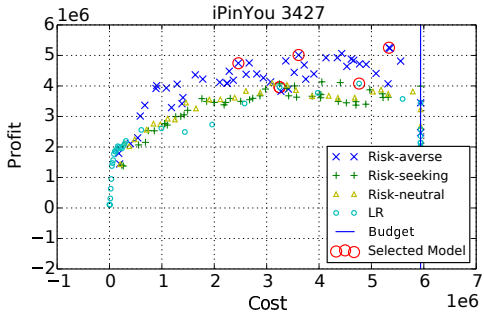}\\
  \vspace{-8pt}
  \caption{Selected model performance comparison with different $\lambda$'s (validation dataset), iPinYou campaign 3427 for RMP strategy with 1/4 budget setting. The models maximizing CP-Profit with $\lambda=0, 0.2, 0.4$ are selected, which is highlighted with red circles. Note that LR and risk-aware models are selected separately and then plotted together.}\label{fig:selected-model-comp-3427-budget}
\end{figure}

\vspace{5pt} \noindent \textbf{Budgeted Settings.}
In practice, advertisers tend to set up campaign budget constraints. With the budget constraint, the basic LR bidding strategy is not necessarily truth-telling, i.e., $\phi \neq 1$ in Eq.~(\ref{eq:lr-bid}) \cite{perlich2012bid}. We performed an analysis on the profit and cost trade-off for different budget constraints. For each campaign, we followed \cite{zhang2015statistical} to use $1/2, 1/4, \ldots, 1/32$ of the original total cost in the test data as the budget. For LR we varied the value of $\phi$ while for VaR and RMP we varied both $\phi$ and $\alpha$ to obtain different profits and costs. As an example, the results of iPinYou campaign 3427 for RMP with 1/4 budget setting are shown in Figure~\ref{fig:selected-model-comp-3427-budget}. We saw that part of the risk-averse strategies dominated other strategies, which was in accordance with the ones without budget constraints.


\vspace{5pt} \noindent \textbf{Distinction with Lowering Bidding.} The performance of risk-averse bidding strategies was not due to lowering bidding. As we know, for a truth-telling linear bidding, if we lower the bid price uniformly, it would result in cost decreasing, profit decreasing and ROI increasing. Unlike this, our risk-averse bidding strategies lowered the bid price of uncertain requests and raised the bid price of confident requests, which led to higher profit and higher ROI with invariant cost. It can be verified in Figure \ref{fig:selected-model-comp-3427-budget}. For example, the left circled LR model lowered the bid price of the right circled LR model by applying a smaller $\phi$, so it achieved lower cost, lower profit and higher ROI. In the meantime, the risk-averse models above the right circled LR model, which had a positive $\alpha$ and a larger $\phi$, achieved the same cost, higher profit and higher ROI.

\subsection{Bidding without Budget Constraints}\label{sec:non-budget-exp}
\begin{figure}[t]
 \centering
 \vspace{-1em}
  \subfigure{\includegraphics[width=0.49\columnwidth]{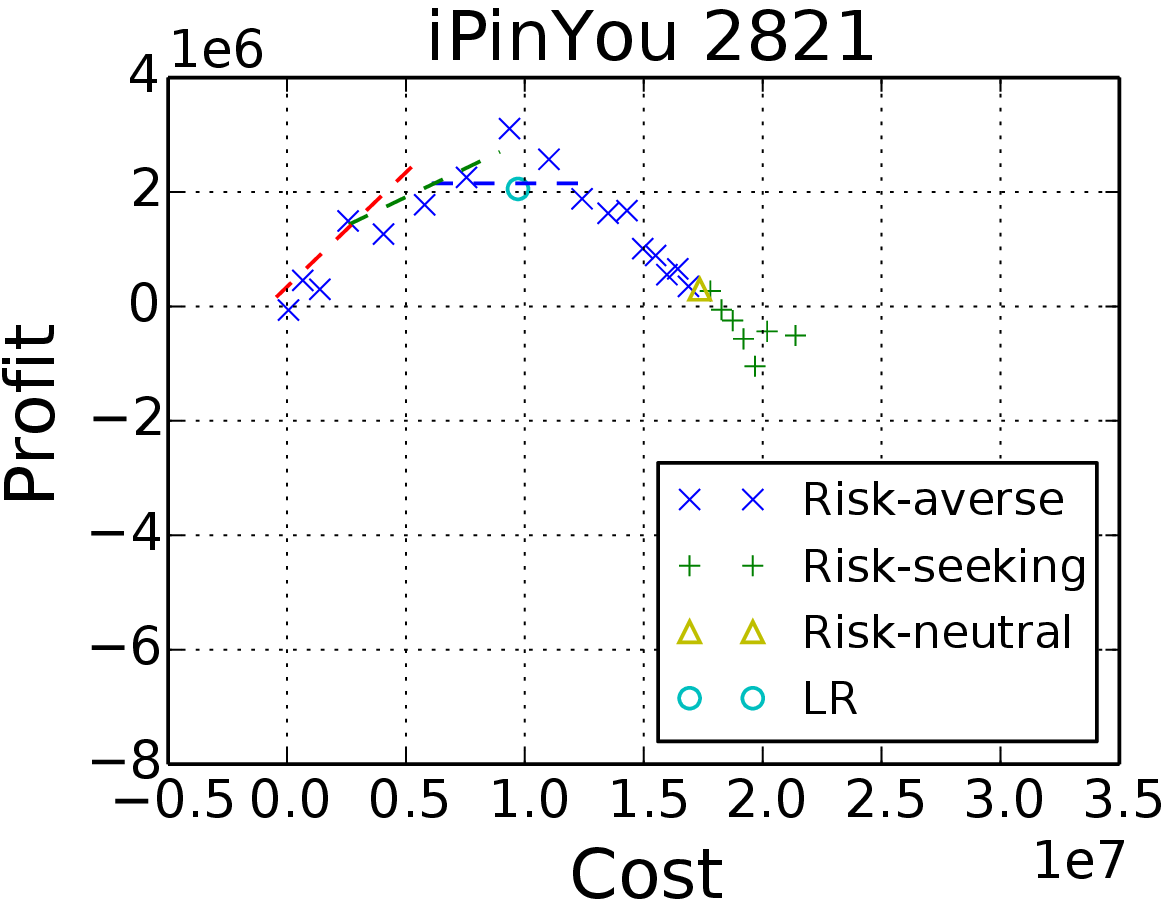}}
  \subfigure{\includegraphics[width=0.49\columnwidth]{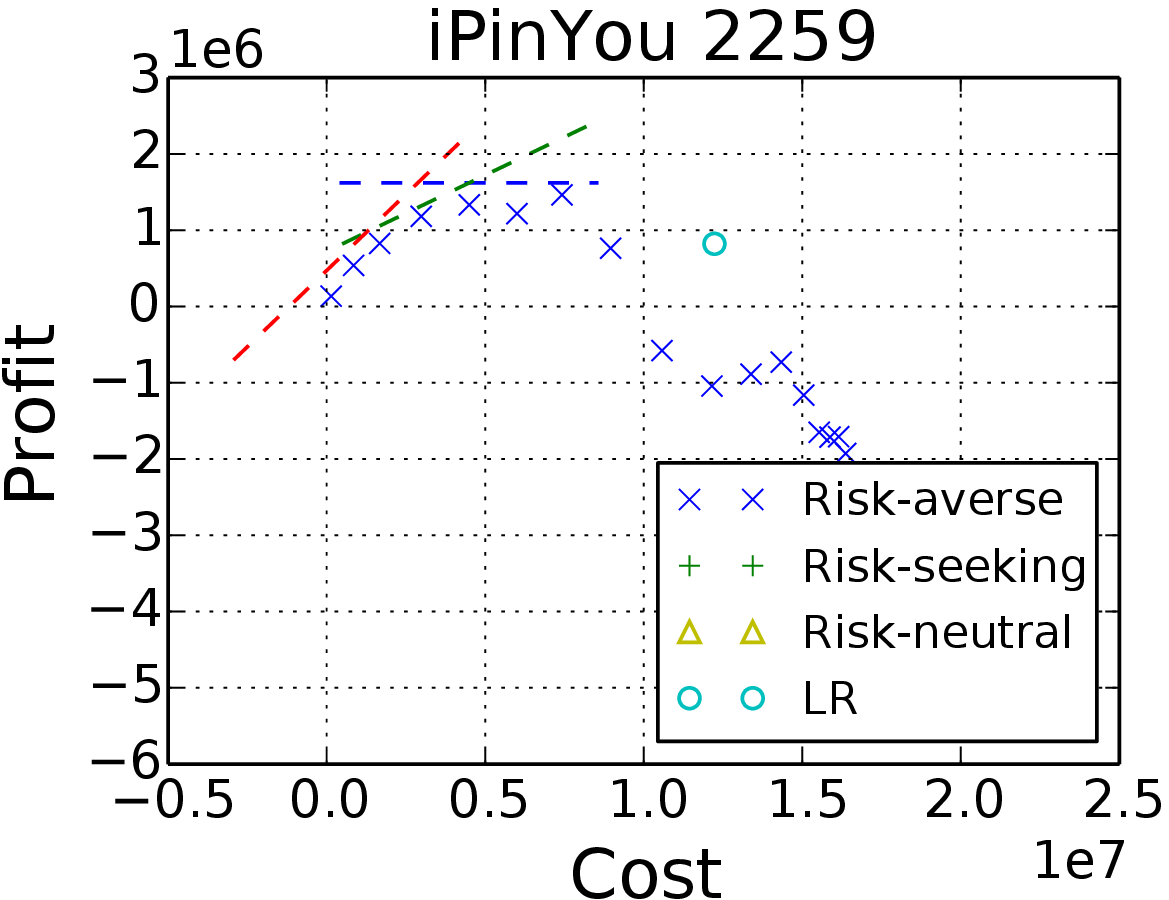}}
 \vspace{-8pt}
 \caption{Model selection. Colored dashed lines stand for $\lambda=0 (\text{blue}), 0.2 (\text{green}), 0.4 (\text{red})$ respectively.}
 \label{fig:rev-cost-analysis-tangent}
\end{figure}

In order to select different risk-aware models (i.e. the strategies' hyperparameter $\alpha$) with various risk-return balanced metrics, we used the proposed CP-Profit metric. Specifically, with a fixed value of $\lambda$, the contour of a specific CP-Profit is a straight line in the cost-profit plot. The highest CP-Profit is achieved by the tangent point of the contour and the cost-profit points. Figure~\ref{fig:rev-cost-analysis-tangent} provides example tangent points with different $\lambda$'s on the cost-profit plot. For a fixed $\lambda$, we then obtained an optimal $\alpha$ that maximizes the CP-Profit.

\begin{figure}[t]
 \centering
 \vspace{-1em}
  \subfigure{\includegraphics[width=0.49\columnwidth]{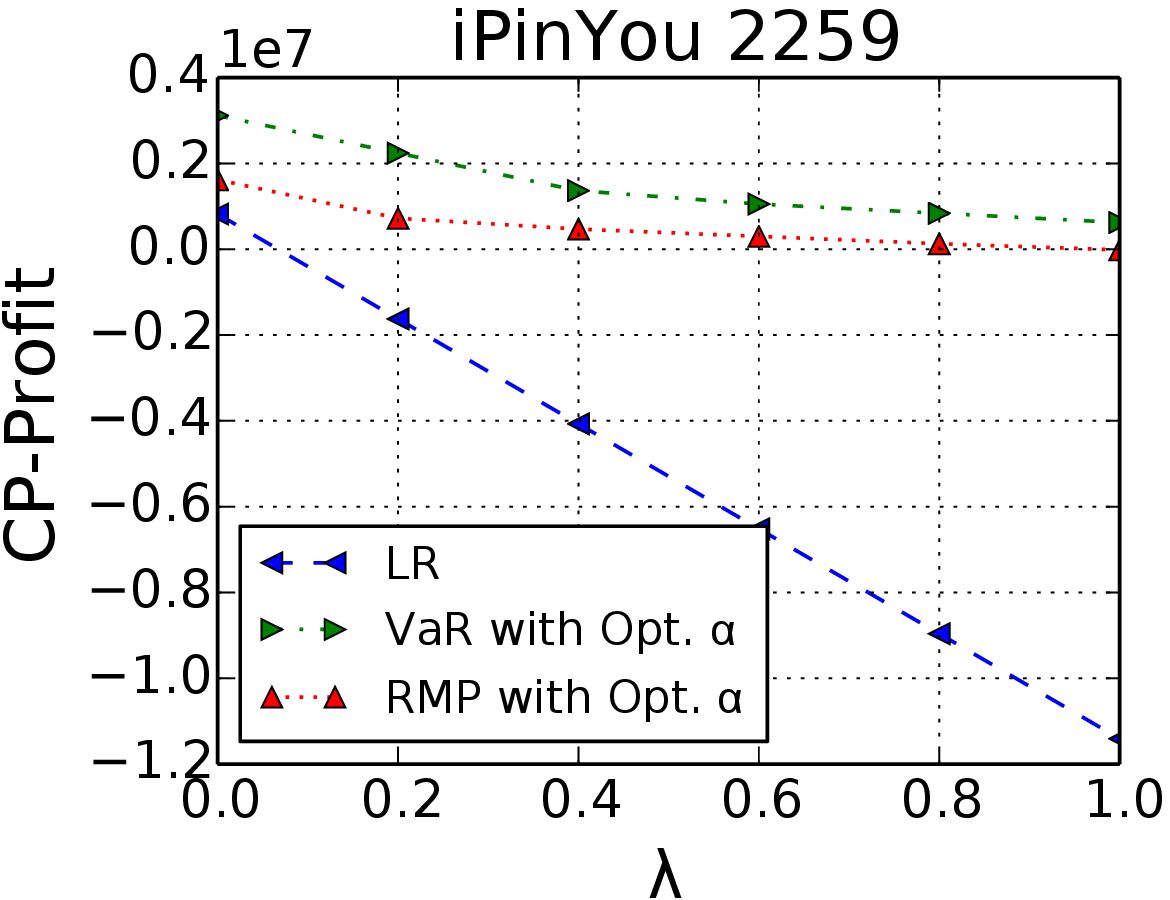}}
  \subfigure{\includegraphics[width=0.49\columnwidth]{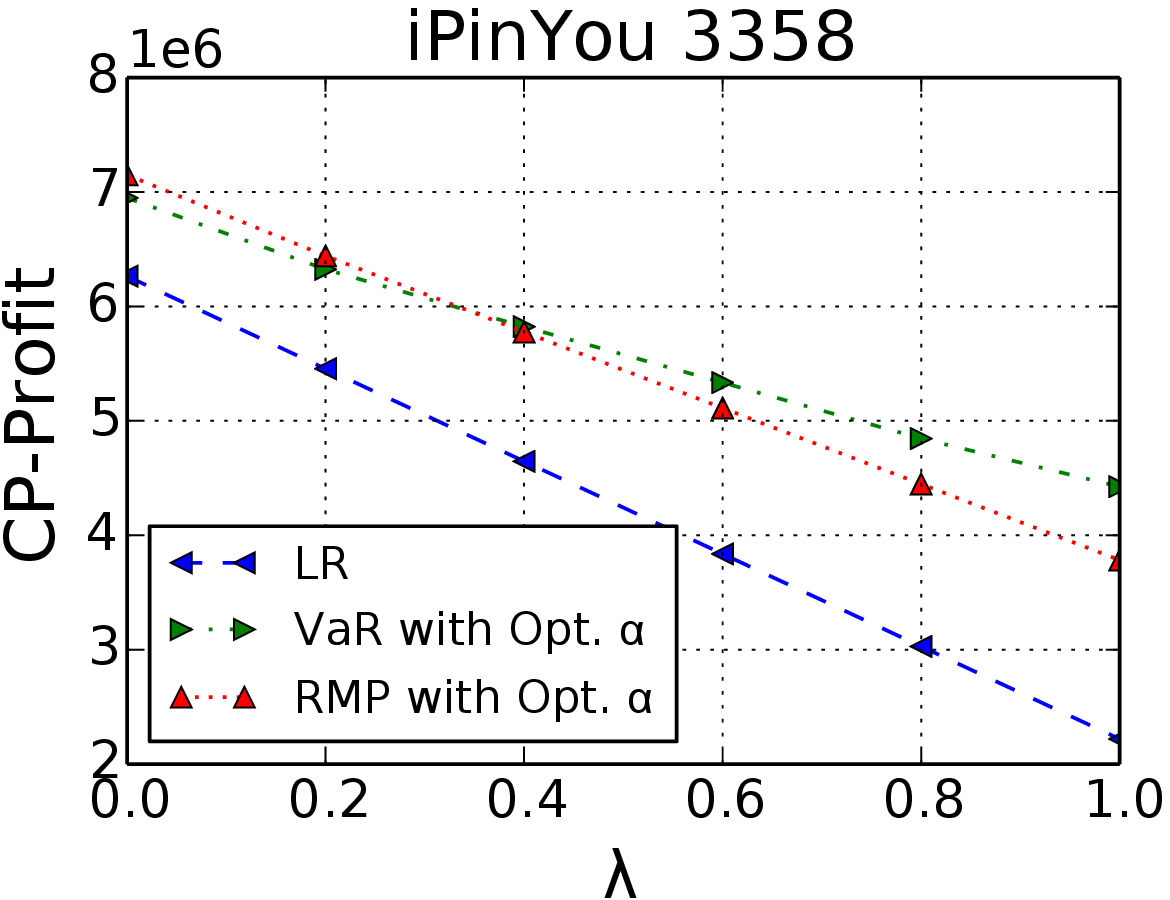}}
 \vspace{-8pt}
 \caption{Selected model performance comparison with different $\lambda$'s (validation dataset).}
 \label{fig:selected-model-comp}
\end{figure}

Figure~\ref{fig:selected-model-comp} plots the performance of the tested bidding strategies against CP-Profit metrics with various $\lambda$ in validation data. We found that: (i) The risk-aware strategies VaR and RMP outperformed the baseline LR on all studied campaigns with all $\lambda$ settings. (ii) Specifically, when $\lambda=0$, i.e., the evaluation metric was purely the profit, VaR and RMP outperformed LR with 50.7\% and 25.1\% improvements. Also note that there were $89\%$ selected models with positive $\alpha$'s, which suggested when there is a fixed volume of bid requests and no limited budget, the risk control helps advertisers make higher profit by spending less money on opportunities with high risk. (iii) For metrics with $\lambda > 0$, the performance gained of VaR and RMP over LR were larger. As we observed that the CP-Profit curves of VaR and RMP were convex while the ones of LR was a straight line, which meant the risk-aware strategies provided higher ROI when the model selection metrics became more conservative, i.e., with higher $\lambda$.


\begin{figure}[t]
  \centering
  \includegraphics[width=1.02\columnwidth]{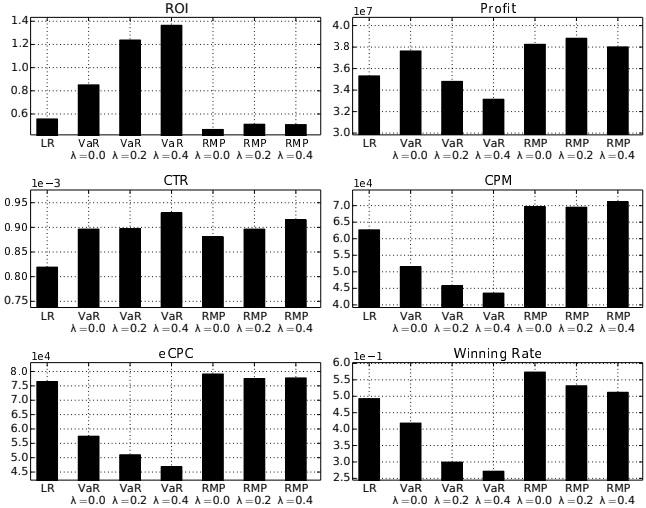}\\
  \vspace{-8pt}
  \caption{Overall non-budgeted test performance.}\label{fig:overall-ipinyou}
\end{figure}

With the model selected on validation data, we compared the strategies on test data. The overall performance across 9 iPinYou campaigns is shown in Figure~\ref{fig:overall-ipinyou}. We observed that:
(i) Most risk-aware bidding strategies yielded significantly higher profit than the baseline LR, which demonstrated the effectiveness of incorporating risk modeling in bid optimization.
(ii) All VaR strategies yielded significantly higher ROI compared to LR, which meant bidding at risk helps advertisers avoid from wasting money on risky opportunities. It turned out that such risky opportunities brought little revenue so avoiding them led to a lower cost and higher profit, and thus much higher ROI.
(iii) RMP strategies stably yielded about 10\% profit more than LR while the ROI was lower than LR. This was because RMP sought the bid price which directly maximized the value at risk of profit, which resulted in consistent profit gains. On the other hand, RMP did not explicitly control the bid price like VaR, thus it did not yield lower CPM or cost. As a result, its ROI was not higher than LR.
(iv) All the risk-aware bidding strategies brought higher CTR than the baseline LR, which might be counterintuitive. Actually, the proposed strategies allocate the budget from the high-uncertainty cases to low-uncertainty ones, which does not mean the average CTR should get lower.

\subsection{Bidding with Budget Constraints}\label{sec:budget-exp}
Following the previous section, we sought the tangent points which maximized CP-Profit with various $\lambda$'s.  The selected models are highlighted with red circles in Figure~\ref{fig:selected-model-comp-3427-budget}.
There might be overlapped red circles as the CP-Profit with different $\lambda$'s still selected the same model.



\begin{figure}[t]
  \centering
  \includegraphics[width=1.02\columnwidth]{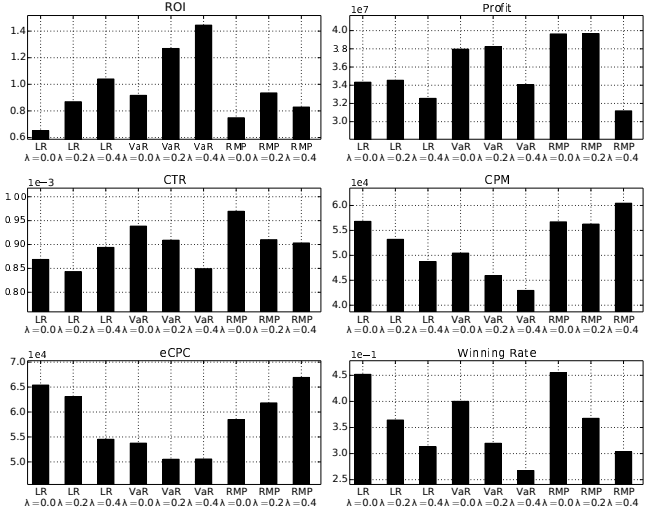}\\
  \vspace{-8pt}
  \caption{Performance with budget constraint (1/2 budget).}\label{fig:overall-1-2-budget-ipinyou}
\end{figure}

Figure~\ref{fig:overall-1-2-budget-ipinyou} provides the overall performance over 9 iPinYou campaigns with budget constraint (1/2 budget setting). The baseline is LR with $\lambda=0$, i.e., simply selecting $\phi$ yielding the highest profit on validation dataset. We found that: (i) For small $\lambda$ settings, VaR and RMP were both better than LR on profit, which verified the claim that the risk-aware bidding strategies successfully yield higher profit via controlling the risk. Specifically, risk-aware bidding strategies performed better on yielding profit for the campaigns with poorer LR estimators. (ii) All proposed methods, i.e., LR with non-zero $\lambda$ model selection and VaR/RMP with various $\lambda$'s, yielded higher ROI than the baseline LR with $\lambda=0$. Some strategies (VaR and RMP with $\lambda=0, 0.2$) yielded both higher profit and higher ROI. (iii) All risk-aware bidding strategies provided higher CTR (except for the conservative VaR with $\lambda=0.4$), which meant the proposed strategies filtered out the low-value cases, which were always with high uncertainty. (iv) For CPM and winning rate, VaR reduced the CPM by lowering bids on unconfident cases and highering the bids on confident ones (via tuning $\phi$); RMP did not guarantee any low bid as it just sought the bid yielding the highest value of risk-controlled profit.

\section{Online Deployment and A/B Test}\label{sec:online}

\begin{figure}
  \centering
  \vspace{-1em}
  \includegraphics[width=1.0\columnwidth]{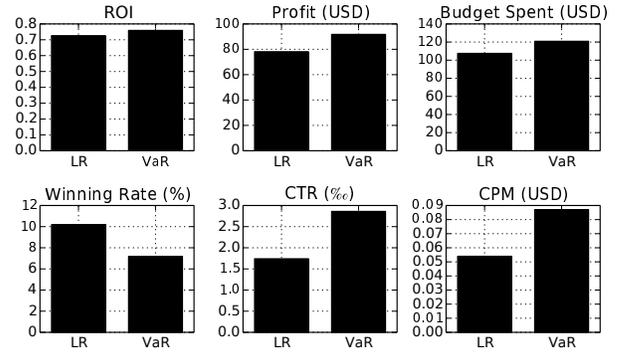}
  \vspace{-2em}
  \caption{Online results from \yoyi~DSP.}\label{fig:online}
\end{figure}

\noindent \textbf{Online Environment.}
We have deployed our bidding strategies on \yoyi~Platform, which is a mainstream DSP in the global RTB market.
The total daily bid request volume received by \yoyi~was about $5$ billion.
Among the received bid requests, about 200M (4\%) met \yoyi's bid target rules of cost-per-click (CPC) campaigns, for which \yoyi~would return a valid bid.
Each of our tested strategies was allocated with $1.5\%$ bid request volume from $10$ CPC campaigns, i.e. about 3M bid opportunities daily.
For training our models, we collected the impression and click data of $7$ consecutive days just before the A/B testing in Jan. 2016, which contained 424M ad impressions, 532K clicks and 47.2K US dollars (USD) expense.
The eCPC on the training data is $0.087$ USD.
A $2\%$ negative instance down sampling was performed by \yoyi~system.
The advertisers of tested CPC campaigns paid a fixed amount for every user landing action, i.e., valid click.
The landing value $v$ was set to be $0.054$ USD, which was the average CPC of all campaigns.

\vspace{5pt}\noindent \textbf{Test Setting.}
We tested two strategies for $7$ days in Jan. $2016$:
(i) The baseline LR strategy. It consists of a logistic regression whose AUC is $88\%$ and a linear bidding function with fixed $\phi=0.56$ set by \yoyi~operation staff.
(ii) The proposed VaR strategy. The parameter $\alpha$ acquired from the training was a positive value that slightly varied daily, which meant the trained bidding strategy risk-averse. The parameter $\phi$ was set to be greater than LR. Thus the two parameters had opposite effects on bid price, which resulted in an average bid price close to that of LR.
Strategies were tested with the same bid volume constraint. The bid volume for each strategy was $19.5$M.
The whole test live volume contained $39$M auctions resulting in $3.3$M impressions and $7.4$K user landings. VaR spent 12\% more budget than LR since $\phi$ was learned from training thus there was no guarantee of equivalent budget spending for A/B test.

\vspace{5pt}\noindent \textbf{Result Discussion.}
The online results are presented in Figure~\ref{fig:online}.
We show the performance over six metrics.
From the results, we found that:
(i) Our VaR strategy outperformed baseline LR on ROI by about 5\%, which demonstrated the efficacy of our risk-aware bidding strategy.
(ii) VaR achieved 17.5\% higher profit than LR. Although this was partly due to the more budget spent by VaR, the more important fact was that VaR strategy could help advertiser spend such more budget at a equivalent or even higher ROI within the same bid request volume. Note that advertisers determine their budget usually depending on ROI and ROI is always negative correlative with budget. Thereby, if a strategy allows an additional amount of budget spent on the same ROI, advertisers would like to increase their budget by that amount.
(iii) VaR won much fewer auctions than LR. According to our setting, VaR bid higher for confident cases and lower for risky cases. Its low winning rate result indicated that the extra confident cases VaR gained were fewer than the risky cases VaR lost.
(iv) As for CTR, VaR achieved 64.4\% higher CTR than baseline LR, which demonstrated the high efficiency of risk-aware strategies in finding high quality ad display opportunities.
(v) VaR got higher CPM than LR. This indicated that our strategy tended to target at impressions with higher quality at the cost of higher price. Although VaR strategy achieved higher CPM, it also achieved higher CTR and the co-effect was reflected in low eCPC and high ROI, which were beneficial.

\vspace{-5pt}
\section{Conclusion}\label{sec:con}
In this paper, we presented a solution for modeling the uncertainty of CTR estimation in RTB display advertising.
With a Bayesian logistic regression CTR estimator, we obtained a closed form of CTR distribution density. On the basis of the distributions of CTR and the market price, two risk-aware bidding strategies are formulated: the first one (VaR) bids the value at risk of the estimated utilities, while the second one (RMP) seeks the optimal bid that maximizes a lower bound of profit with a controlled risk. Our risk-return analysis and offline experiment demonstrated 15.4\% profit gain over a linear optimized bidding strategy. To test the applicability of our risk-aware bidding strategies in a live setting, the bidding strategies have been deployed on an operational platform. 17.5\% profit gain over the baseline was observed in a 7-day online A/B test.


For future work, we will analyze the market equilibrium if more advertisers adopt the risk-aware bidding strategies. We also plan to further explore  applications from the proposed CTR distribution model and the risk-aware bid optimization framework. For an ad placement with multi-frame dynamic creatives, a portfolio optimization \cite{elton2009modern} can be performed to select the optimal creatives combination to yield the highest profit with a controlled risk. Moreover, our CTR distribution model is potential to be dynamically updated to evaluate users' interest on an advertised product after repeated ad displays over time.

\vspace{5pt}\noindent \textbf{Acknowledgement} This work is financially supported by National Natural Science Foundation of China (61632017).

\vspace{-2pt}
{\small
\bibliographystyle{abbrv}
\bibliography{wsdm282-zhang}
}

\section{APPENDIX}


\vspace{5pt} \noindent \textbf{VaR Realization.}
Our VaR strategy relies on the CTR distribution model from Section~\ref{sec:ctr-model}.
The parameters $\mu_i$ and $q_i$ are trained offline, while $\sum_i \mu_i x_i$ and $\sum_i q_i^{-1} x_i$ for a given bid request is calculated online, which only takes double time compared to a traditional LR model.
We then calculate $\mathbb{E}[\hat{y}]$ and $\std[\hat{y}]$ from $\sum_i \mu_i x_i$ and $\sum_i q_i^{-1} x_i$, which is, however, quite costly.
Fortunately, we can move this calculation from online to offline by building up a look-up table, with $(\sum_i \mu_i x_i, \sum_i q_i^{-1} x_i)$ as its key and $(\mathbb{E}[\hat{y}], \std[\hat{y}])$ as its value.
Since $\sum_i \mu_i x_i$ and $\sum_i q_i^{-1} x_i$ are in particular ranges respectively, we can discretize the value with enough accuracy and calculate $\mathbb{E}[\hat{y}]$ and $\std[\hat{y}]$ of the discrete $(\sum_i \mu_i x_i, \sum_i q_i^{-1} x_i)$ combinations.
For each bid request online, we round ($\sum_i \mu_i x_i$, $\sum_i q_i^{-1} x_i$) to get the discrete key and then read $\mathbb{E}[\hat{y}]$ and $\std[\hat{y}]$ from the look-up table with $O(1)$ time.

\vspace{5pt} \noindent \textbf{VaR Efficiency.} For offline look-up table building, if we separate $\sum_i \mu_i x_i$ and $\sum_i q_i^{-1} x_i$ into 1000 bins respectively and sample $1000$ times from the corresponding $\hat{y}$ distribution, it needs $10^{9}$ sampling operations in total, which can be done in a few minutes on a modern PC. For online bidding, as discussed above, it has equal computational complexity with logistic regression.

\vspace{5pt} \noindent \textbf{RMP Realization.} We solve Eq.~(\ref{eq:bid-rmr}) using a look-up table too, which takes $(\sum_i \mu_i x_i, \sum_i q_i^{-1} x_i)$ as its key and bid price $b$ as its value. For building the look-up table, we offline enumerate $b$ within a limited range. For each enumerated $b$, we sample market price $z$ and estimated CTR $\hat{y}$ according to their distributions for sufficient times and calculate $R(b)$ for each sample. After that we obtain the objective $\mathbb{E}[R(b)] - \alpha \std[R(b)]$ for each $b$. Finally we choose the optimal $b$ that maximizes the objective and store the solution in the look-up table. When online bidding, we just calculate $\sum_i \mu_i x_i$ and $\sum_i q_i^{-1} x_i$ for the bid request, then find the optimal $b$ from the look-up table with $O(1)$ time.

\vspace{5pt} \noindent \textbf{RMP Efficiency.} Compared to VaR strategy, RMP additionally enumerates $b$ and samples $z$ in offline calculation. In practice, $1000$ enumerations are enough as the range of bid price is quite small. Note that, for each key of the look-up table, all the enumerated bid prices $b$ can share the same samples of $\hat{y}$ and $z$. Thus, the computational complexity of offline calculation for RMP only has constant difference with the one for VaR, which takes a few minutes as discussed above. When online bidding, RMP strategy has equal computational complexity with logistic regression, which makes it capable in real-world business.

\vspace{5pt} \noindent \textbf{Analysis of RMP strategy.}
To give an insight into solving $b$ in Eq.~(\ref{eq:bid-rmr}), we analyze the relationship between $\mathbb{E}[R(b)]$ and $\std[R(b)]$. Given a bid $b$, the expectation of $R(b)$ is formulated as
\begin{align}
\mathbb{E}[R(b)] = v \cdot \mathbb{E}[\hat{y}] \cdot z_0 - z_1,
\end{align}
where $z_k$ denotes $\int_0^b z^k p_z(z) dz$ for simplicity. And the expected squared profit is
\begin{align}
\mathbb{E}[R(b)^2] = v^2 \mathbb{E}[\hat{y}^2] z_0 - 2 v \mathbb{E}[\hat{y}] z_1 + z_2.
\end{align}

Based on $\mathbb{E}[R(b)]$ and $\mathbb{E}[R(b)^2]$, the variance of profit is
\begin{align}
\var[R(b)] = v^2 \mathbb{E}[\hat{y}^2] z_0 - 2 v \mathbb{E}[\hat{y}] z_1 + z_2 - (v \mathbb{E}[\hat{y}] z_0 - z_1)^2
\end{align}

We further analyze the change trends of the expectation $\mathbb{E}[R(b)]$ and variance $\var[R(b)]$ w.r.t. the bid price $b$.
\begin{align}
\frac{\partial \mathbb{E}[R(b)]}{\partial b} = p_z(b) \cdot (v \mathbb{E}[\hat{y}] - b), \label{eq:rmr-e-derivative}
\end{align}
which is positive when $b < v \mathbb{E}[\hat{y}]$ and negative otherwise.

We are also interested in the change trend of $\var[R(b)]$ in the range around the truth-telling bid price. We obtain
\begin{align}
\frac{\partial \var[R(b)]}{\partial b} \bigg|_{b=v \mathbb{E}[\hat{y}]} = p_z(v \mathbb{E}[\hat{y}]) \cdot v^2 \cdot \var[\hat{y}] \geq 0. \label{eq:rmr-var-derivative}
\end{align}

\begin{figure}[t]
  \centering
  \includegraphics[width=.70\columnwidth]{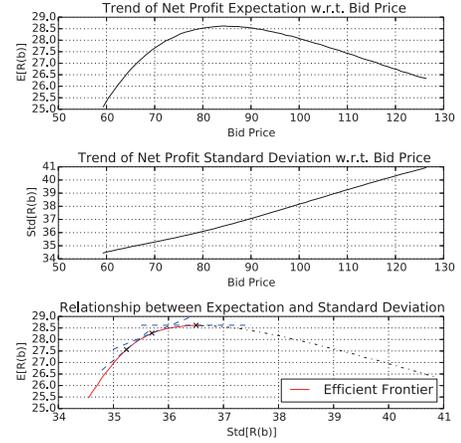}\\
  \vspace{-3pt}
  \caption{An example of $\mathbb{E}[R(b)]$, $\std[R(b)]$ trend w.r.t. bid price, and the relationship between them.}\label{fig:profit-expectation-std-trend}
\end{figure}

Therefore, the variance increases in a range around the conventional truth-telling bid price $v \mathbb{E}[\hat{y}]$, so does the standard deviation. Figure~\ref{fig:profit-expectation-std-trend} shows an example of the relationship between $\mathbb{E}[R(b)]$ and $\std[R(b)]$, where the parameters of CTR distribution and market price distribution are the same as Figure~\ref{fig:ctr-mp-rev-pdf}.

Based on the reasonable preference of higher $\mathbb{E}[R(b)]$ and lower $\std[R(b)]$, we further borrow the concept of \emph{\text{efficient} frontier} from finance \cite{hull2012risk}.
Every point at the efficient frontier corresponds to the optimal bid price defined by Eq.~(\ref{eq:bid-rmr}) with a particular $\alpha$. For example, the slopes $\alpha$ of the dashed lines are $0,1,2$ respectively. The tangent points are of the maximum $\mathbb{E}[R(b)] - \alpha \std[R(b)]$, each of which corresponds to a particular bid price, i.e., the solution of Eq.~(\ref{eq:bid-rmr}).

\end{document}